\RequirePackage{fix-cm}
\documentclass[twocolumn]{svjour3}          
\smartqed  
\usepackage{graphicx}
\usepackage{amssymb,amsmath}
\usepackage{comment}
\usepackage[misc]{ifsym}
\RequirePackage{xspace}
\newcommand{\uin}{\ensuremath{U_{\text{in}}(\boldsymbol{x})}\xspace}
\newcommand{\uvar}{\ensuremath{U(\boldsymbol{\theta})}\xspace}
\newcommand{\partheta}{\ensuremath{\boldsymbol{\theta}}\xspace}
\newcommand{\uphi}{\ensuremath{U_{\phi}(\boldsymbol{x})}\xspace}
\newcommand{\niter}{\ensuremath{N_{\text{iter}}}\xspace}
\newcommand{\npar}{\ensuremath{N_{\text{par}}}\xspace}
\newcommand{\nvar}{\ensuremath{N_{\text{var}}}\xspace}
\newcommand{\nevttr}{\ensuremath{N_{\text{event}}^{\text{train}}}\xspace}
\newcommand{\nevtts}{\ensuremath{N_{\text{event}}^{\text{test}}}\xspace}
\newcommand{\ndepthin}{\ensuremath{N_{\text{in}}^{\text{depth}}}\xspace}
\newcommand{\ndepthvar}{\ensuremath{N_{\text{var}}^{\text{depth}}}\xspace}

\journalname{Computing and Software for Big Science}
\begin{document}

\title{Event Classification with Quantum Machine Learning in High-Energy Physics}
\subtitle{}

\author{
Koji Terashi \and
Michiru Kaneda \and
Tomoe Kishimoto \and
Masahiko Saito \and
Ryu Sawada \and
Junichi Tanaka
}

\institute{K.~Terashi~(\Letter) \and M.~Kaneda \and T.~Kishimoto \and M.~Saito \and R.~Sawada \and J.~Tanaka \at 
              7-3-1 Hongo, Bunkyo-ku, Tokyo 113-0033, Japan\\
              International Center for Elementary Particle Physics (ICEPP), The University of Tokyo\\
              \email{koji.terashi@cern.ch} (corresponding author)
}

\date{Received: date / Accepted: date}

\maketitle

\begin{abstract}

We present studies of quantum algorithms exploiting machine learning 
to classify events of interest from background events, 
one of the most representative machine learning applications in high-energy physics.
We focus on variational quantum approach to learn the properties of input data 
and evaluate the performance of the event classification using both simulators and 
quantum computing devices. Comparison of the performance 
with standard multi-variate classification techniques based on a boosted-decision tree
and a deep neural network using classical computers
shows that the quantum algorithm has comparable performance 
with the standard techniques at the considered ranges of the number of input variables and
the size of training samples.
The variational quantum algorithm is tested with quantum computers,
demonstrating that the discrimination of interesting events from background is feasible. 
Characteristic behaviors observed during a learning process using quantum circuits
with extended gate structures are discussed, as well as 
the implications of the current performance to the application 
in high-energy physics experiments.

\end{abstract}

\section{Introduction}
\label{sec:intro}
The field of particle physics has been recently driven by large experiments to collect and analyze data produced in particle collisions
occurred using high-energy accelerators. In high-energy physics (HEP) experiments, particles created by collisions are 
observed by layers of high-precision detectors surrounding collision points, producing a large amount of data. 
The large data volume has motivated the use of 
machine learning (ML) techniques in many aspects of experiments 
to improve their performances 
(see, e.g, \cite{hepmllivingreview} for a living review of ML techniques to particle physics).
In addition, computational resources are expected to be reduced for specific tasks 
by adopting relatively new techniques such as ML.
This will continue over next decades; for example, a next-generation proton-proton collider, 
called High-Luminosity Large Hadron Collider (HL-LHC)~\cite{ApollinariG.:2017ojx,Evans:2008zzb}, 
at CERN~\footnote{The European Organization for Nuclear Research located in Geneva, Switzerland, https:://www.cern.ch} 
is expected to deliver a few exabytes of data every year and 
requires very large computing resources for the data processing.
Quantum computing (QC), on the other hand, has been evolving rapidly over the past years, with a promise of a significant
speed-up or reduction of computational resources in certain tasks. Early attempts to use QC for HEP have been made, 
e.g, on data analysis~\cite{Mott:2017xdb,Zlokapa:2019lvv}, 
identification of charged particle trajectories~\cite{Shapoval:2019txi,Bapst:2019llh,Zlokapa:2019tkn,Tuysuz:2020ocw}, 
reconstruction of particle collision points~\cite{Das:2019hrw} and particle spray 
called jets~\cite{Wei:2019rqy}, as well as the simulation of event generation 
called parton shower~\cite{Bauer:2019qx,Bauer:2019qxa}. 
The techniques developed in HEP are also adapted to QC, e.g, 
the unfolding techniques for physics measurement are applied to QC in Refs.~\cite{Cormier:2019kcq,Bauer:2019uf}.
Among these attempts, the quantum machine learning (QML) is considered as one of the 
QC algorithms that could bring quantum advantages over classical methods, 
as discussed in literatures, e.g,~\cite{Preskill2018quantumcomputingin}. 

Most frequently-used ML technique in HEP data analysis is the discrimination of 
events of interest, e.g, signal events originating from new physics beyond the Standard Model of 
particle physics, from background events. 
In this paper, we have investigated the application of QML techniques to the task of the event classification in HEP data analysis.
To our knowledge, the first attempt to utilize QC for HEP data analysis is performed in Ref.~\cite{Mott:2017xdb} for 
the classification of the Higgs boson using quantum annealing~\cite{Johnson2011Quantum}.

\begin{figure}
\includegraphics[width=0.49\textwidth]{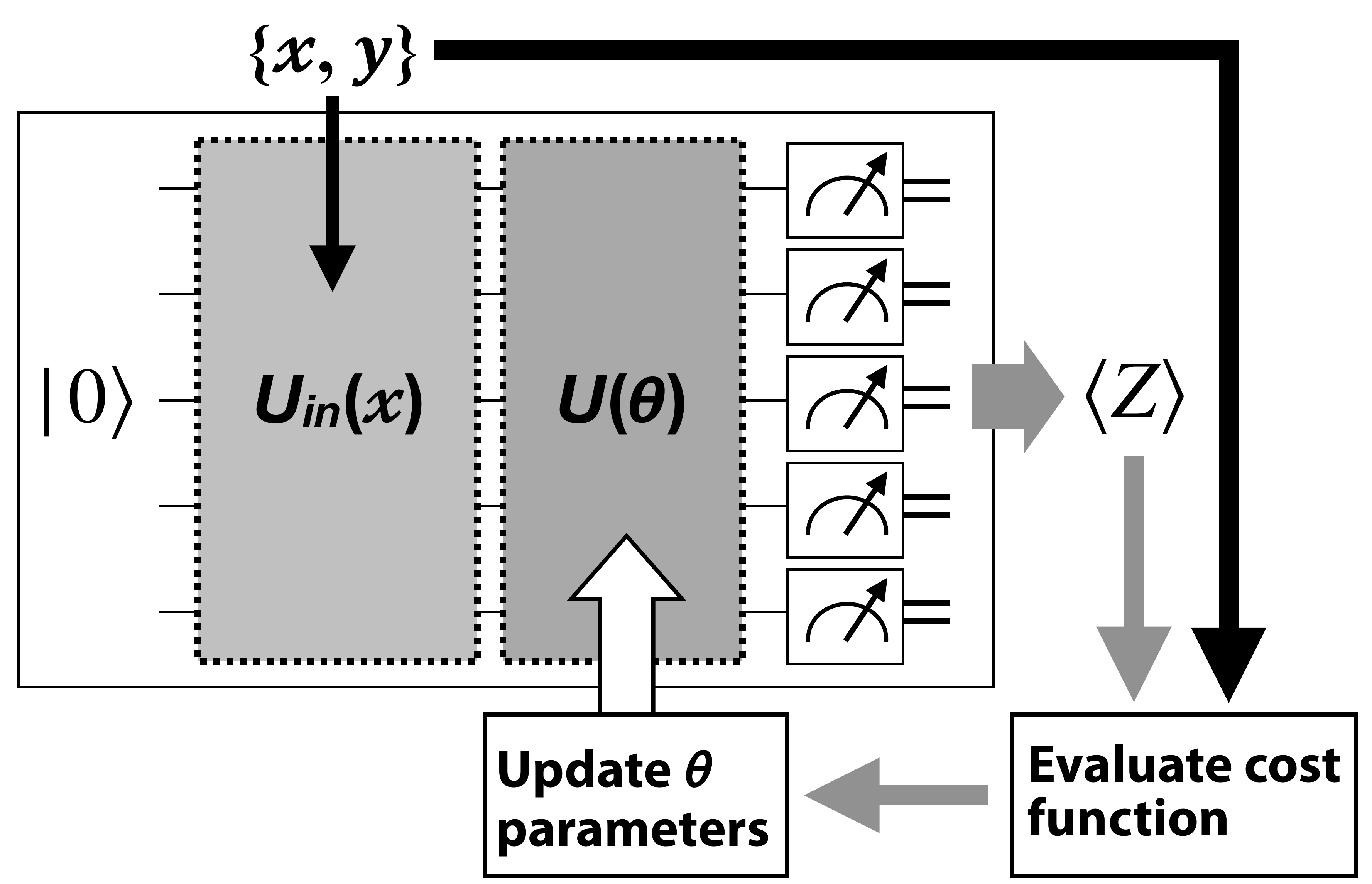}
\caption{Representative variational quantum circuit used for the event classification in this study.
$\{\boldsymbol{x},\boldsymbol{y}\}$ is a set of pairs composed of an input data $\boldsymbol{x}$ and 
an input label $\boldsymbol{y}$~(desired output value).
$\langle Z\rangle$ is an output from the quantum circuit. 
The components of the circuit and their roles are explained in the text.}
\label{fig:var_circuit} 
\end{figure}

We focus on QML algorithms developed for gate-based quantum computer, in particular the algorithms
based on variational quantum circuit~\cite{Peruzzo2014vqe}. 
In the variational circuit approach, the classical input data are encoded into quantum states and a quantum computer is used to obtain and measure the quantum states which vary with tunable parameters. 
Exploiting a complex Hilbert space that grows exponentially with the number of ``quantum bits" (or qubits) in
quantum computer, the representational ability of the QML is far superior to classical ML that 
grows only linearly with the number of classical bits. 
This motivates the application of ML techniques to quantum computer, which could lead to an advantage 
over the classical approach.
The optimization of the parameters is performed using classical computer, 
therefore the variational method is considered to
be suitable for the present quantum computer, which has difficulty in processing deep quantum circuits 
due to limited quantum coherence.
Practically, actual performance of the variational quantum algorithm depends on
the implementation of the algorithm and the properties of the QC device.
The primary aim of this paper is to demonstrate the feasibility of ML for the event classification in HEP data analysis
using gate-based quantum computer. 

First, the variational quantum algorithms are described in Sect.~\ref{sec:algo}, followed by the classical approaches 
that are used for the comparison. Section~\ref{sec:setup} discusses the experimental setup used in the study, including 
the dataset, software simulator and quantum computer. Results of the experiments are discussed 
in Sect.~\ref{sec:result}, followed by discussions on several observations about the performance of 
the quantum algorithms in Sect.~\ref{sec:discussion}. We conclude the studies in Sect.~\ref{sec:conclusion}.

\section{Algorithms}
\label{sec:algo}
\subsection{Variational Quantum Approaches}
\label{subsec:qc_algo}

In this study we consider an approach based on variational quantum circuit with tunable parameters~\cite{Peruzzo2014vqe}. 
The quantum circuit used in this algorithm is constr-ucted, 
as shown in Fig.~\ref{fig:var_circuit}, using three components: 1) quantum gates to encode 
classical input data $\boldsymbol{x}$ into quantum states (denoted as \uin),  
2) quantum gates to produce output states used for supervised learning (denoted as \uvar)
and 3) measurement gates to obtain output values from the circuit, that 
are subsequently compared with the corresponding input labels $\boldsymbol{y}$. 
In this study, the measurement is performed 1,024 times on each event with the Pauli-$Z$ operators
and the average value of the measurements is used for improving the statistical accuracy. 
For the classification of events into two categories, the first two qubits are typically measured.
The \uvar gates used in 2) are parameterized 
such that they are optimized to model input training data by iterating the computational processes of 1)--3) 
by \niter times and tuning the parameters \partheta. 
The parameter tuning is performed using a classical computer by minimizing a cost function, 
which is defined such that a difference
between the input labels $\boldsymbol{y}$ and the measured values $\langle Z\rangle$ can be quantified.
The optimized \uvar circuit with the tuned parameters is used, with the same \uin gates, 
to classify unseen data for testing.
The \uin and \uvar are often built by using a same set of quantum gates with different parameters multiple times 
to enhance the representational ability of the data.
The numbers of the repetition used for the  \uin and \uvar are denoted by \ndepthin and \ndepthvar, respectively.

In this study, we use two implementations of the 
variational quantum algorithms, called Quantum Circuit Learning (QCL)~\cite{Mitarai_2018} and 
Variational Quantum Classification (VQC)~\cite{Havlcek2019SupervisedLW}.
The QCL is used for testing the performance of the variational quantum algorithm on simulator.
The VQC is used for testing the algorithm on both real quantum computer and simulator with 
small samples.

\begin{figure}
\includegraphics[width=0.5\textwidth]{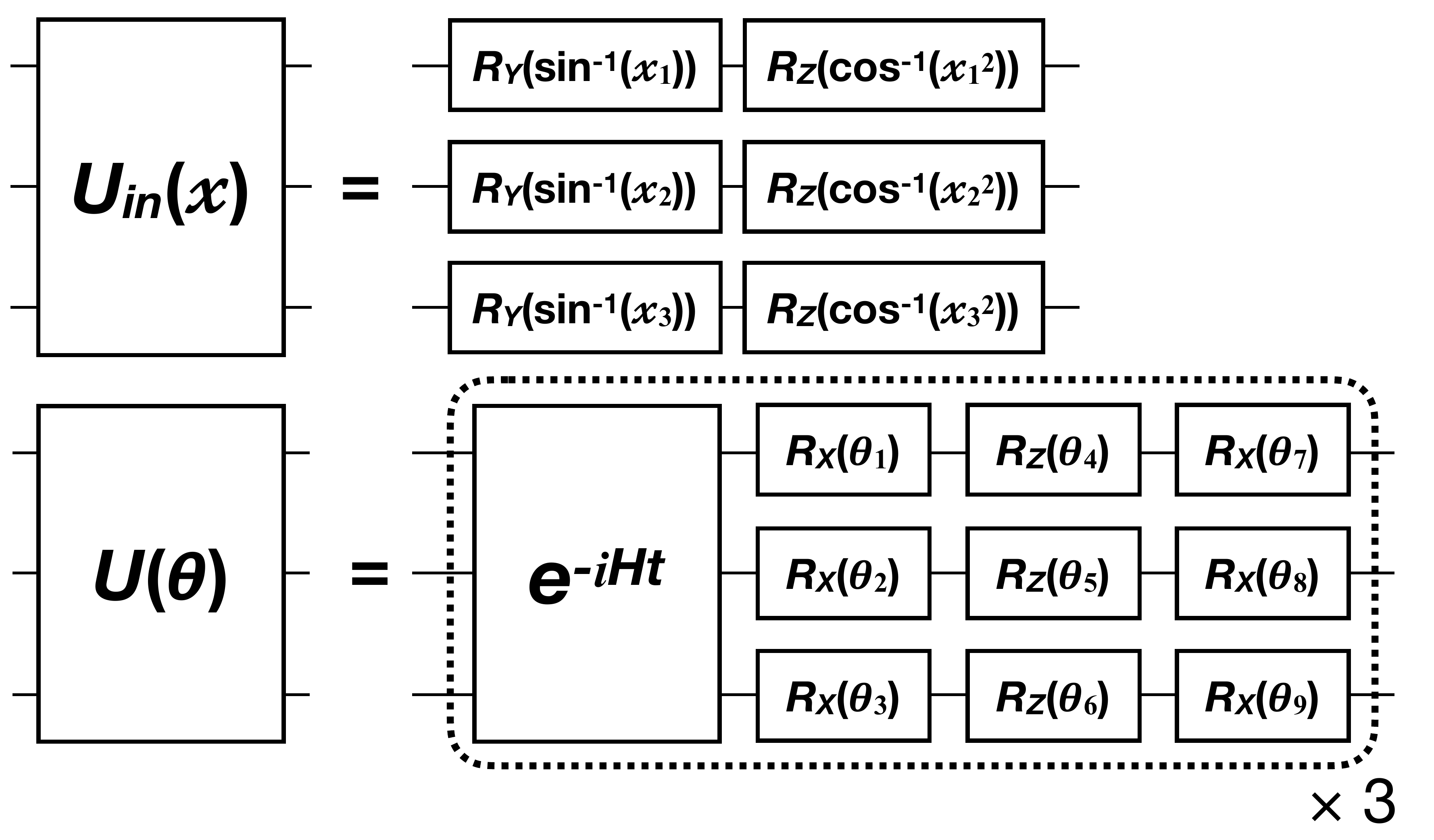}
\caption{\uin and \uvar circuits used in this study for the QCL algorithm.}
\label{fig:qcl_circuit} 
\end{figure}

\subsubsection{Quantum Circuit Learning}

A QCL circuit used in this study for the 3-variable classification is shown in Fig.~\ref{fig:qcl_circuit}. 
The \uin in QCL is characterized by the series of single-qubit rotation gates $R_Y$ and $R_Z$~\cite{Mitarai_2018}.
The angles of the rotation gates are obtained from the input data $\boldsymbol{x}$ to be $\sin^{-1}(\boldsymbol{x})$ and $\cos^{-1}(\boldsymbol{x}^2)$, respectively. 
The input data are needed to be normalized within the range [$-1$, $1$] by scaling linearly
using the maximum and minimum values of the input variables.
The normalization is performed separately for the training and testing samples to avoid data 
beyond the [$-1$, $1$] range. In this case, the classification performance is slightly suboptimal 
for the testing sample. The effect is however checked to be small by comparing the performance 
with the case where the testing sample is normalized with the scaling derived 
from the training sample and clipped to [$-1$, $1$].
The \uvar is constructed using a time-evolution gate, denoted as $e^{-iHt}$,  
with the Hamiltonian $H$ of an Ising model with random coefficients 
(for creating entanglement between qubits)
and the series of $R_X$, $R_Z$ and $R_X$ gates with angles as parameters. 
The nominal \ndepthvar value is set to 3 after optimization studies. 
This results in 27 parameters in total for the 3-variable case. 
The structure for the 5- and 7-variable circuits is the same as the 3-variable case,
leading to the total parameters of 45 and 63, respectively. 
The measurement is performed on the first two qubits using the Pauli-$Z$ operators,
and the outcome of the measurement is fed into the cost function via softmax. 
The cost function is defined using a cross-entropy function in scikit-learn package~\cite{scikit-learn},
and the minimization of the cost function is performed using COBYLA.
See \cite{Mitarai_2018} for more details about the implementation.

\subsubsection{Variational Quantum Classification}

\begin{figure}
\includegraphics[width=0.5\textwidth]{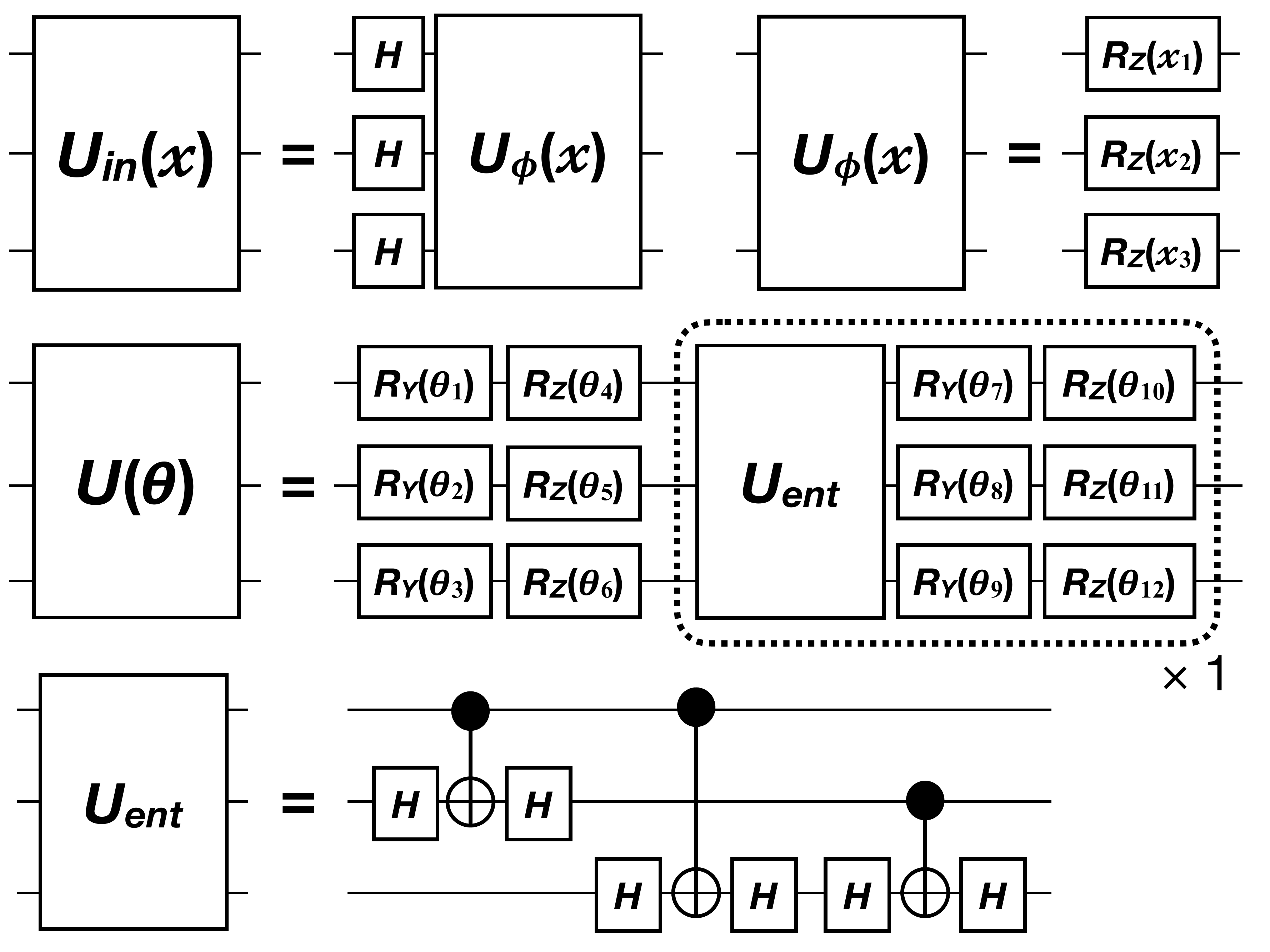}
\caption{\uin and \uvar circuits used in this study for the VQC algorithm.}
\label{fig:vqc_circuit} 
\end{figure}

Figure~\ref{fig:vqc_circuit} shows a VQC circuit for the 3-variable classification used in this study.
The \uin consists of a set of Hadamard gates and rotation gates with angles from the input data $\boldsymbol{x}$ 
(the latter is represented as \uphi in the figure). 
The \uphi is composed of single-qubit rotation gates of the form 
$U_{\phi_{\{k\}}}(\boldsymbol{x})=\exp{(i\phi_{\{k\}}(\boldsymbol{x})Z_k)}$, a diagonal phase gate 
with the linear function of $\phi_{\{k\}}(\boldsymbol{x})=x_k$. This is identical to the one used in 
Ref.~\cite{Havlcek2019SupervisedLW} as the single-qubit gate (see Eq.~(32) of the supplementary information 
of Ref.~\cite{Havlcek2019SupervisedLW}), and is referred to as the``First Order Expansion" (FOE).
The \uphi is not repeated in this study unless otherwise stated, thus $\ndepthin=1$.
The \uvar part of the circuit is also taken from that in \cite{Havlcek2019SupervisedLW} but simplified by not repeating a set of entangling gate 
($U_\text{ent}$) and single-qubit rotation gates $R_Y$ and $R_Z$ (surrounded by the dashed box in Fig.~\ref{fig:vqc_circuit}). 
The $U_\text{ent}$ is implemented using the Hadamard and CNOT gates, as in Fig.~\ref{fig:vqc_circuit}.
The total number of \partheta parameters is 12 (20, 28) for the 3 (5, 7)-variable classification.
The measurement is performed on all the qubits using the Pauli $Z$ operators, and the measured outcomes are fed into
the cost function. 
The cost function for the VQC algorithm is a cross-entropy function and the minimization 
 is performed using COBYLA as well.

\subsection{Classical Approaches}
\label{subsec:classical_algo}

The ML application to the classification of events has been widely attempted in HEP data analyses.
Among others, a Boosted Decision Tree (BDT) in the TMVA framework~\cite{Speckmayer_2010} 
is one of the most commonly used algorithms.
A neural network (NN) is another class of multivariate analysis methods, and 
an algorithm with a deep neural network (DNN) has been proven to be powerful for modelling complex 
multi-dimensional problems.
We use BDT and DNN as benchmark tools for comparison with the performance of the variational quantum algorithms.

In this study we use the TMVA package~4.2.1 for the BDT and the Keras~2.1.6 with TensorFlow~1.8.0 backend 
for the DNN.
The BDT and DNN parameters used are summarized in Table~\ref{tab:bdt_dnn_param}. 
The maximum depth of the decision tree~(MaxDepth) and the number of trees in the forest~(NTrees) vary 
with the number of events used in the training (\nevttr) to avoid over-training.
The DNN model is a fully-connected feed-forward network composed of 2--6 hidden layers with 16--256 nodes each.
The numbers of hidden layers and nodes per layer are varied according to \nevttr to avoid over-training.

\begin{table}
\caption{Parameter settings for the BDT and DNN used in this study. The definitions of the BDT parameters are 
documented in Ref.~\cite{Speckmayer_2010}.}
\label{tab:bdt_dnn_param}
\begin{tabular}{ll}
\hline\noalign{\smallskip}
BDT Parameter & Value  \\
\noalign{\smallskip}\hline\noalign{\smallskip}
BoostType &  Grad\\
NTrees & 10 ($\nevttr=0.1\text{K}$),\\ 
& 100 ($0.5\text{K}\leq\nevttr\leq10\text{K}$),\\ 
& 1000 ($\nevttr\geq50\text{K}$)\\ 
MaxDepth & 1 ($\nevttr\leq1\text{K}$),\\ 
& 2 ($5\text{K}\leq\nevttr\leq100\text{K}$),\\
& 3 ($\nevttr\geq200\text{K}$)\\
nCuts & 20\\
MinimumNodeSize & 2.5\%\\
UseBaggedBoost & True\\
BaggedSampleFraction & 0.5\\
\noalign{\smallskip}\hline
\end{tabular}
\begin{tabular}{ll}
\hline\noalign{\smallskip}
DNN Parameter & Value  \\
\noalign{\smallskip}\hline\noalign{\smallskip}
Layer Type &  Dense\\
Number of hidden layers & 2 ($\nevttr=0.1\text{K}$ or 1K),\\ 
& 3 ($\nevttr=0.5\text{K}$),\\ 
& 4 ($5\text{K}\leq\nevttr\leq100\text{K}$),\\ 
& 6 ($\nevttr\geq200\text{K}$)\\ 
Number of nodes per & 16 ($0.1\text{K}\leq\nevttr\leq0.5\text{K}$),\\ 
hidden layer & 64 ($1\text{K}\leq\nevttr\leq10\text{K}$),\\ 
& 128 ($50\text{K}\leq\nevttr\leq100\text{K}$),\\ 
& 256 ($\nevttr\geq200\text{K}$)\\ 
Activation function & rectified linear unit\\
Optimizer & Adam\\
Learning rate & 0.001\\
Batch size & None ($\nevttr\leq10\text{K}$),\\
& 2048 ($\nevttr\geq50\text{K}$)\\
Batch normalization & No\\
Number of epochs & 100 with early stopping\\
\noalign{\smallskip}\hline
\end{tabular}
\end{table}

\section{Experimental Setup}
\label{sec:setup}
Our experimental test of the variational quantum algorithms is performed using both simulators of quantum computers and real quantum computers available via the IBM Q Network~\cite{IBMQ}. 
As a benchmark scenario for the HEP data analysis, we consider a problem of discriminating events with 
Supersymmetry (SUSY) particles from the most representative background events.

\subsection{Dataset}
\label{subsec:data}
We use the ``SUSY Data Set" available in the UC Irvine Machine Learning Repository~\cite{Dua:2019}, which was prepared for studies of Ref.~\cite{Baldi:2014kfa}.
The signal process, labelled true, targets a chargino-pair production via a Higgs boson. 
Each chargino decays into a neutralino that escapes detection and a $W$-boson 
that subsequently decays into a charged lepton and a neutrino, resulting in a final state 
with two charged leptons and a missing transverse momentum.
The background process, labelled false, is a $W$-boson pair production ($WW$) with each $W$-boson 
decaying into a charged lepton and a neutrino. 
Therefore, both the signal and background processes have the same final state.
Monte Carlo simulation is used to produce events of these processes as described in \cite{Baldi:2014kfa}.

In our main studies a small fraction of the data is used
because the process of the full data~(5 million events) with the quantum algorithms requires significant computing resources.
For the comparison of the quantum and classical MLs,
five sets of data containing 100, 500, 1,000, 5,000 and 10,000 events are used for training and
other five sets of data with the same number of events for testing.
For the classical MLs, additional four sets of data containing 
50,000, 100,000, 200,000 and 500,000 events are used to study the dependence on the sample size.

\begin{figure*}
\includegraphics[width=1.01\textwidth]{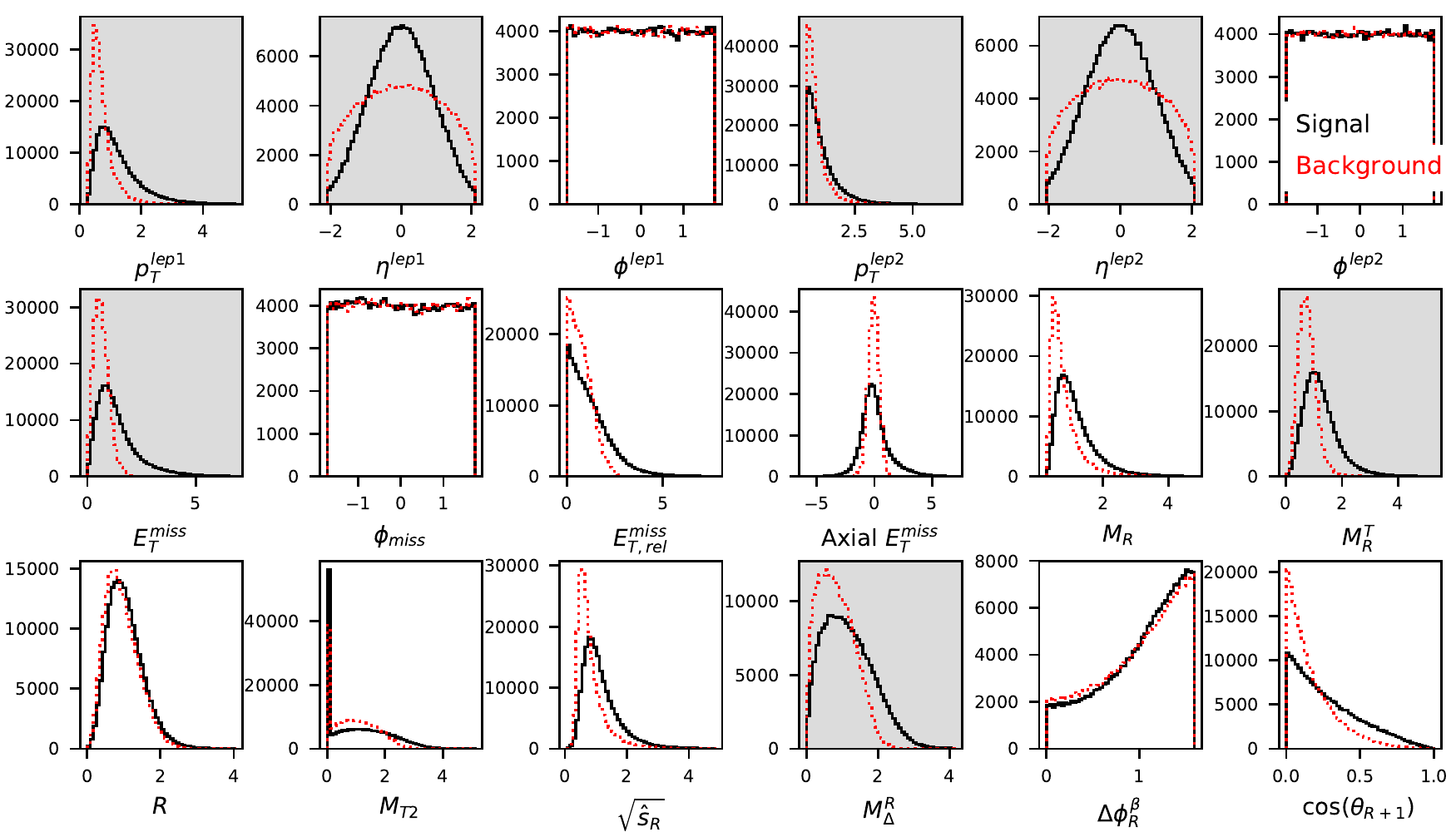}
\caption{Normalized distributions of input variables for SUSY signal (solid histograms) and background (dashed histograms). 
The first nine variables up to $E_{\text{T, rel}}^{\text{miss}}$ are low-level features and the last nine are high-level ones. 
The main variables used in the study are highlighted in grey on the background while all the variables are considered
for the DNN and BDT, as discussed in the text.}
\label{fig:SUSY_input_vars}
\end{figure*}

The dataset contains 18 variables characterizing the properties of the SUSY signal and $WW$ background events, 
ranging from low-level variables such as lepton transverse momenta to high-level variables 
such as those reflecting the kinematics of $W$-bosons and/or charginos (detailed in \cite{Baldi:2014kfa}).
Figure~\ref{fig:SUSY_input_vars} shows the normalized distributions of the 18 variables for 
the signal and background events. 
Among those, the following 3, 5 and 7 variables, which are quoted as $\nvar=3$, 5 and 7 later, are considered in the main study:
\begin{description}
  \item[\hspace*{3mm}3 variables: ] $p_\text{T}^{\text{lep1}}$, $p_\text{T}^{\text{lep2}}$ and $E_T^\text{miss}$,
  \item[\hspace*{3mm}5 variables: ] 3 variables above, $M_{R}^{T}$, $M_{\Delta}^{R}$,
  \item[\hspace*{3mm}7 variables: ] 5 variables above, $\eta^{\text{lep1}}$, $\eta^{\text{lep2}}$.
\end{description}
The choice of these variables is made as follows: first, the combination of 3 variables is determined by testing 
different combinations of the variables using the DNN algorithm and taking the one with the highest 
AUC~(area under ROC curve) value. Starting with the selected variables of $\nvar=3$, 
more variables are sequentially added and determined in the same way for $\nvar=5$ and 7.
In addition, all the 18 variables are used for evaluating the best performance
which the classical MLs can reach (Sect.~\ref{subsec:sim_result}).

\subsection{Simulator}
\label{subsec:sim}
We use quantum circuit simulators to evaluate the performance of the quantum algorithms.
The QCL circuit is implemented using Qulacs~0.1.8~\cite{Qulacs}, 
a fast quantum circuit simulator implemented in C++, 
with Python 3.6.5 and gcc~7.3.0, and the performance is evaluated on cloud Linux servers managed by 
OpenStack at CERN. The Qulacs supports the use of GPU, but it is not exploited in this study.
The VQC circuit is implemented using Aqua~0.6.1 in the Qiskit~0.14.0~\cite{Qiskit}, a quantum computing 
software development framework (Qiskit Aq-ua framework).
The VQC performance is evaluated using a QASM simulator on a local machine
as well as real quantum computer explained in the next section. 
No wall-time comparison is made between the simulators in this study. 

The Qulacs simulator has capability of executing the variational quantum algorithm with more variables or 
more data events than the QASM simulator. This allows us to evaluate the performance of 
the quantum algorithm in more realistic settings. 

\subsection{Quantum Computer}
\label{subsec:qc}
We use the 20-qubit IBM Q Network quantum computers, called Johannesburg~\cite{johannesburg} and 
Boeblingen~\cite{boeblingen}, for evaluating the VQC performance.
The quantum computers are accessed using the {\it QuantumInstance} class in the Qiskit Aqua framework. 
The \uin part of the VQC circuit (Fig.~\ref{fig:vqc_circuit}) is created separately for each event 
because the  \uphi gates depend on the input data $\boldsymbol{x}$. 
For the training and testing, we use 40 events each, composed of 20 signal and 20 background events. 
The \partheta parameters are determined by iterating the training process as explained in Sect.~\ref{subsec:qc_algo}.
The \niter is set to 100 unless otherwise stated.

\section{Results}
\label{sec:result}
\subsection{Qulacs Simulator}
\label{subsec:sim_result}

First, the classification performance of the QCL algorithm evaluated using the Qulacs simulator 
is compared with those of the BDT and DNN. 
Due to a significant increase of the computational resources with \nvar for the QCL 
(discussed in Sect.~\ref{subsec:QCL_usage}), the \nvar is considered only up to 7.

\begin{figure}
\includegraphics[width=0.53\textwidth]{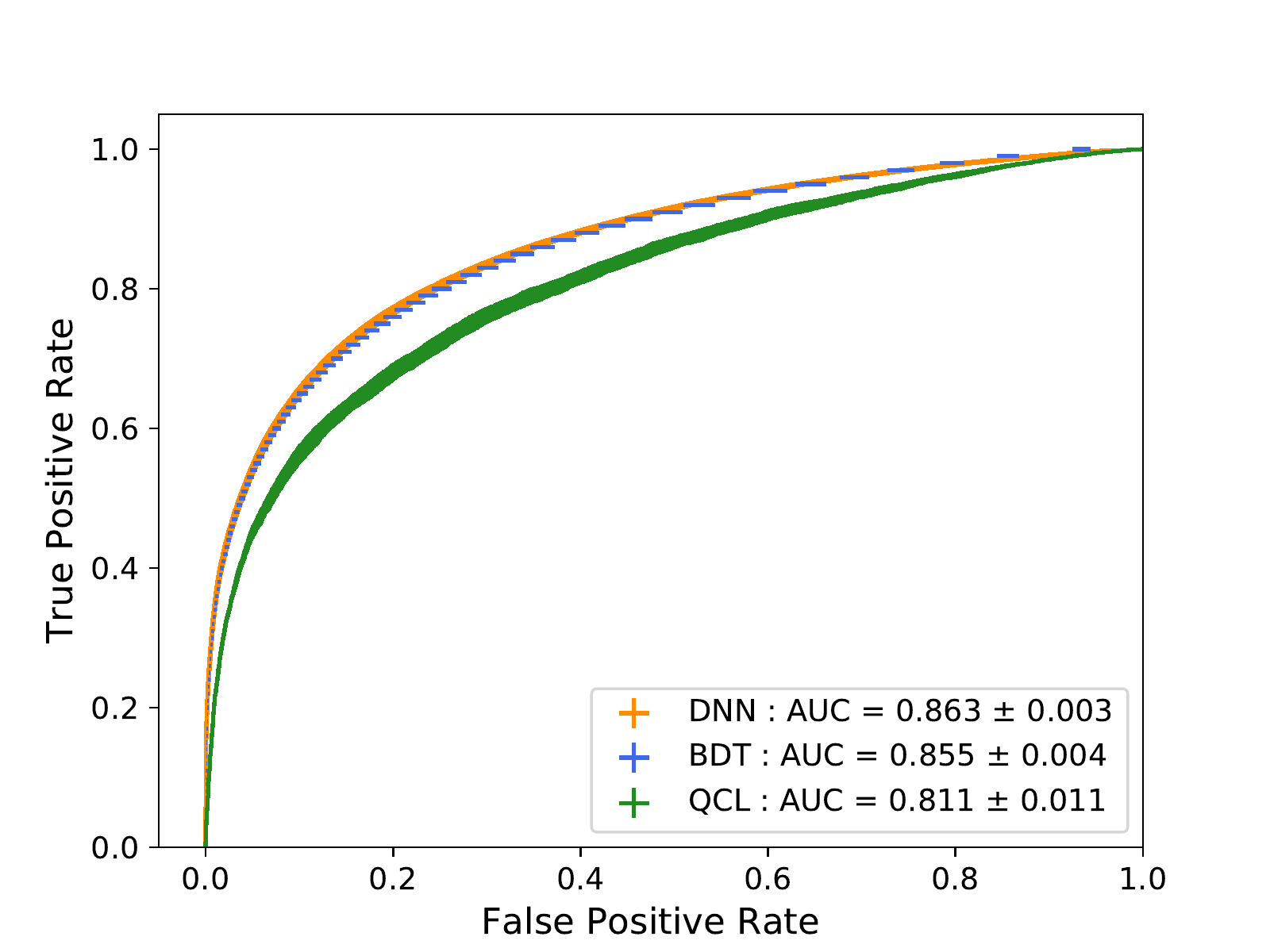}
\caption{ROC curves obtained from the test sample for the BDT, DNN and QCL algorithms 
with $\nvar=7$ and $\nevttr=10,000$. 
The error bands correspond to the standard deviations of the values obtained by repeating the 
calculation over the training and test samples.}
\label{fig:result_roc} 
\end{figure}

\begin{figure}
\includegraphics[width=0.53\textwidth]{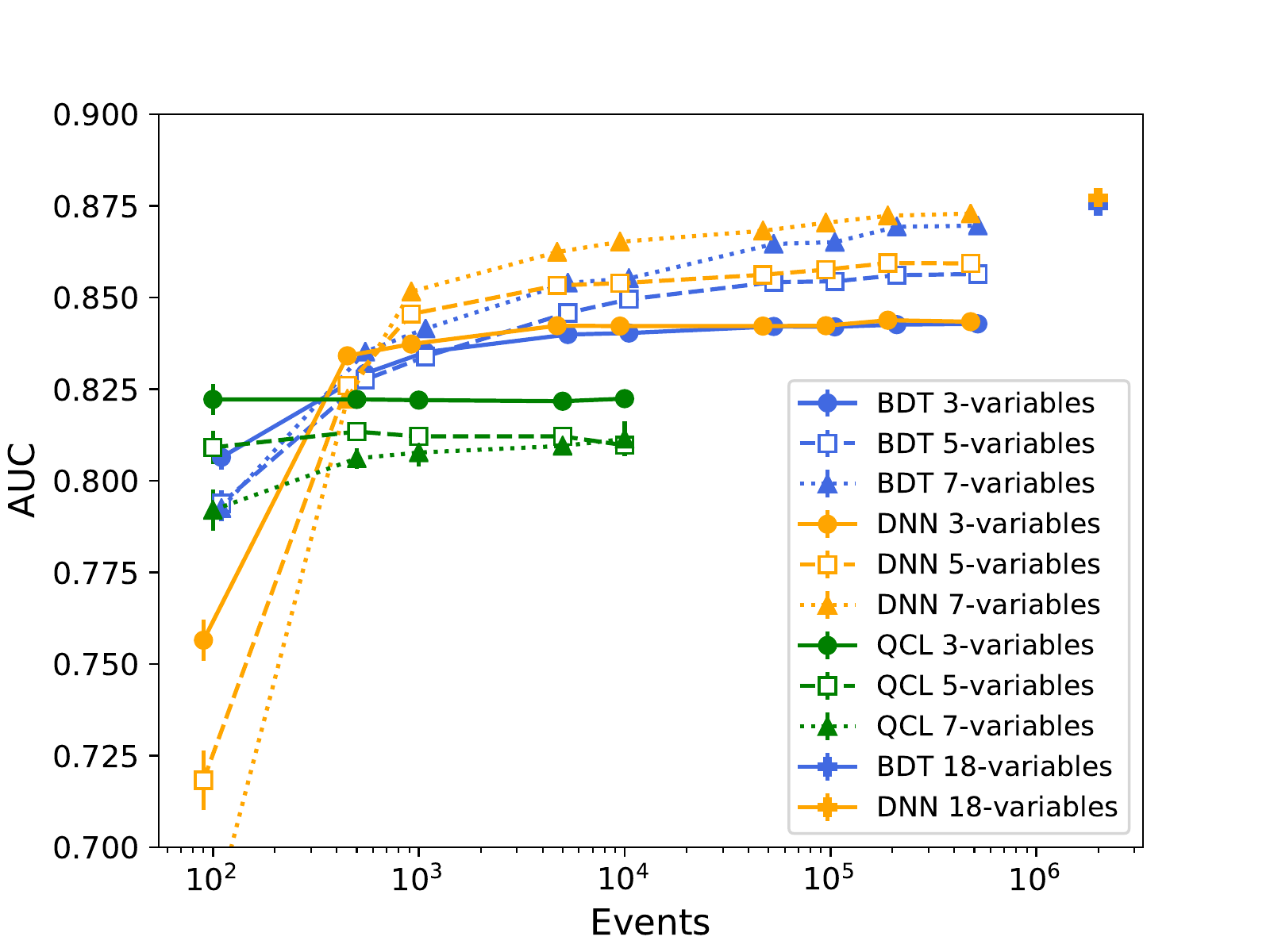}
\caption{Average AUC values (calculated from the test samples) as a function of the training sample size 
for the BDT, DNN and QCL algorithms with $\nvar=3$ (circles), 5 (squares) and 7 (triangles).
For the BDT and DNN, the average AUC values for the training sample of 2,000,000 events and 18 variables 
are also shown with the plus markers.
The error bars represent the standard deviations of the average AUC values. The BDT and DNN points are 
slightly shifted horizontally from the nominal \nevttr values to avoid overlapping.}
\label{fig:result_auc_all} 
\end{figure}

Figure~\ref{fig:result_roc} shows ROC curves in the testing for the three algorithms 
with $\nvar=7$ and $\nevttr=10,000$, and Figure~\ref{fig:result_auc_all} shows the comparisons of the AUC values 
as a function of \nevttr for $\nvar=3$, 5 and 7.
For each algorithm, a single AUC value is obtained from a test sample after each training, and the calculation is 
repeated 100 (30) times at $\nevttr\leq10,000$ ($50,000\leq\nevttr\leq500,000$). 
The center and the width of each curve in Fig.~\ref{fig:result_roc} correspond to the average value and 
the standard deviation of the true/false positive rates obtained from the repeated calculations 
over the training and test samples. 
Shown in Fig.~\ref{fig:result_auc_all} is the average of the resulting AUC values
and the standard deviations of the average.
As expected, it is apparent from the BDT and DNN curves that the performance of these two algorithms improves rapidly 
with increasing \nevttr and then flattens out. 
The BDT works well over the entire \nevttr range 
while the DNN performance appears to improve faster at very small \nevttr
and exceed BDT at \nevttr beyond $\sim1,000$. 
In the case of $\nvar=7$ and $\nevttr=500,000$, the AUC values are $0.8729\pm0.0003$ for the DNN and 
$0.8696\pm0.0006$ for the BDT. 
When using all the 18 variables with 2,000,000 events for the training and testing each, 
the average AUC value from only five trials is $0.8772\pm0.0004$ ($0.8750\pm0.0004$) for the DNN (BDT).

The performance of the QCL algorithm is characterized by the relatively flat AUC values regardless of \nevttr.
Increasing the \nvar appears to degrade the performance if the \nevttr is fixed, 
and the same behavior is also seen for the DNN with $\nevttr\leq500$ (not clearly visible for the BDT). 
Further studies show that the QCL performance of the flat AUC value and the degradation with increasing \nvar
is related to the choice of the variables: the $\nvar=3$ variables used have sufficient information 
for the QCL algorithm to discriminate the signal from background, and no positive impact is seen on the performance
by adding more variables or more events. However, it is seen that the performance improves by adding them 
if different combinations of the variables are selected.
The DNN algorithm overcomes the degradation and eventually improves the performance with increasing \nvar  
by using more data. 
Investigating how the QCL algorithm behaves with more data is a future subject.
Nevertheless, for the \nvar and \nevttr ($\leq$ 10,000) ranges considered 
all the three algorithms have a comparable discriminating power with the AUC values of 0.80--0.85.

\begin{figure}
\includegraphics[width=0.53\textwidth]{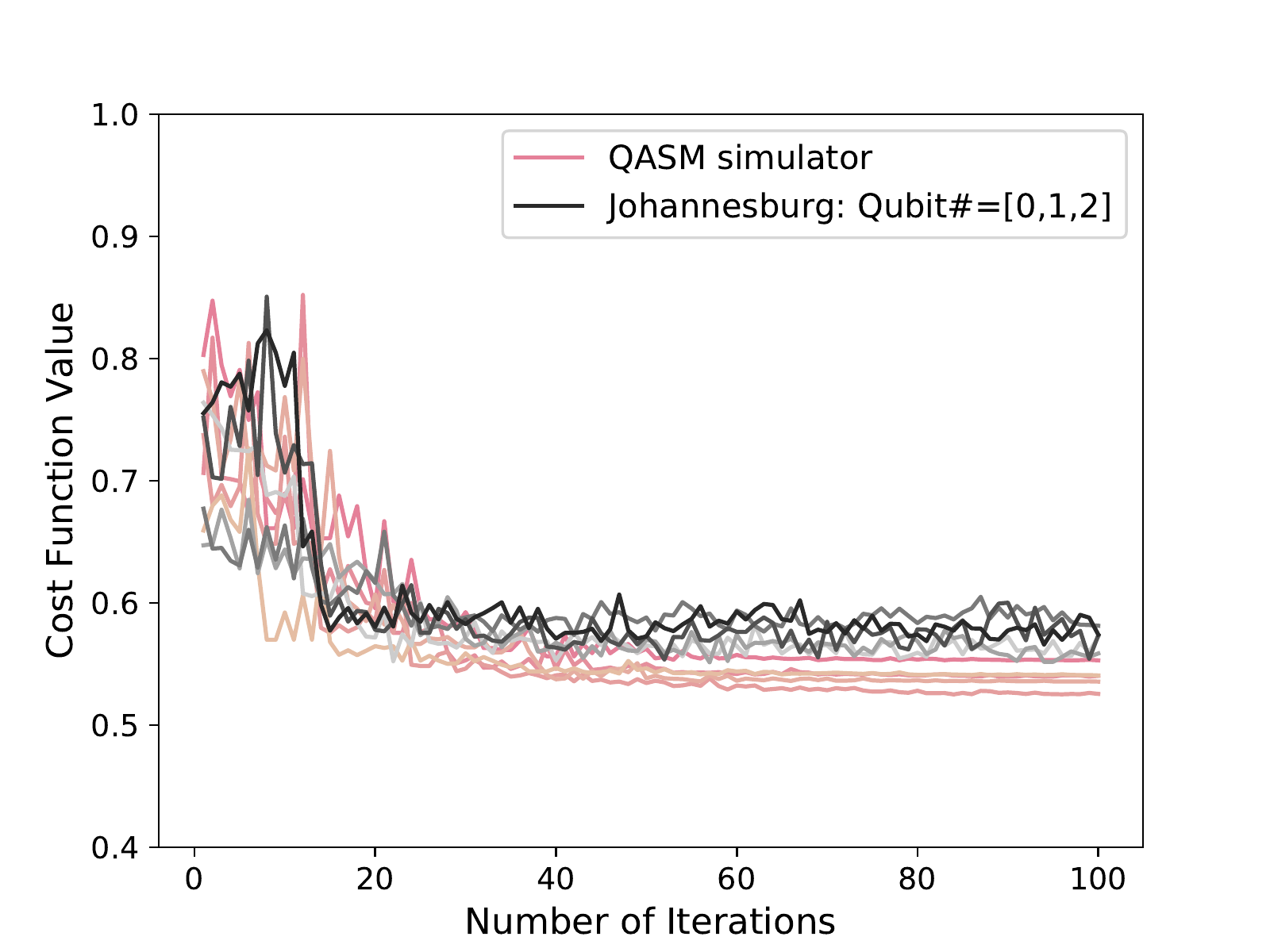}
\caption{Evolution of the cost function value in the training of the VQC algorithm with $\nvar=3$ and $\nevttr=40$.
Shown are the cost function values observed in 5 training trials for quantum computer and QASM simulator.}
\label{fig:cost_vqc} 
\end{figure}

\begin{figure}
\includegraphics[width=0.53\textwidth]{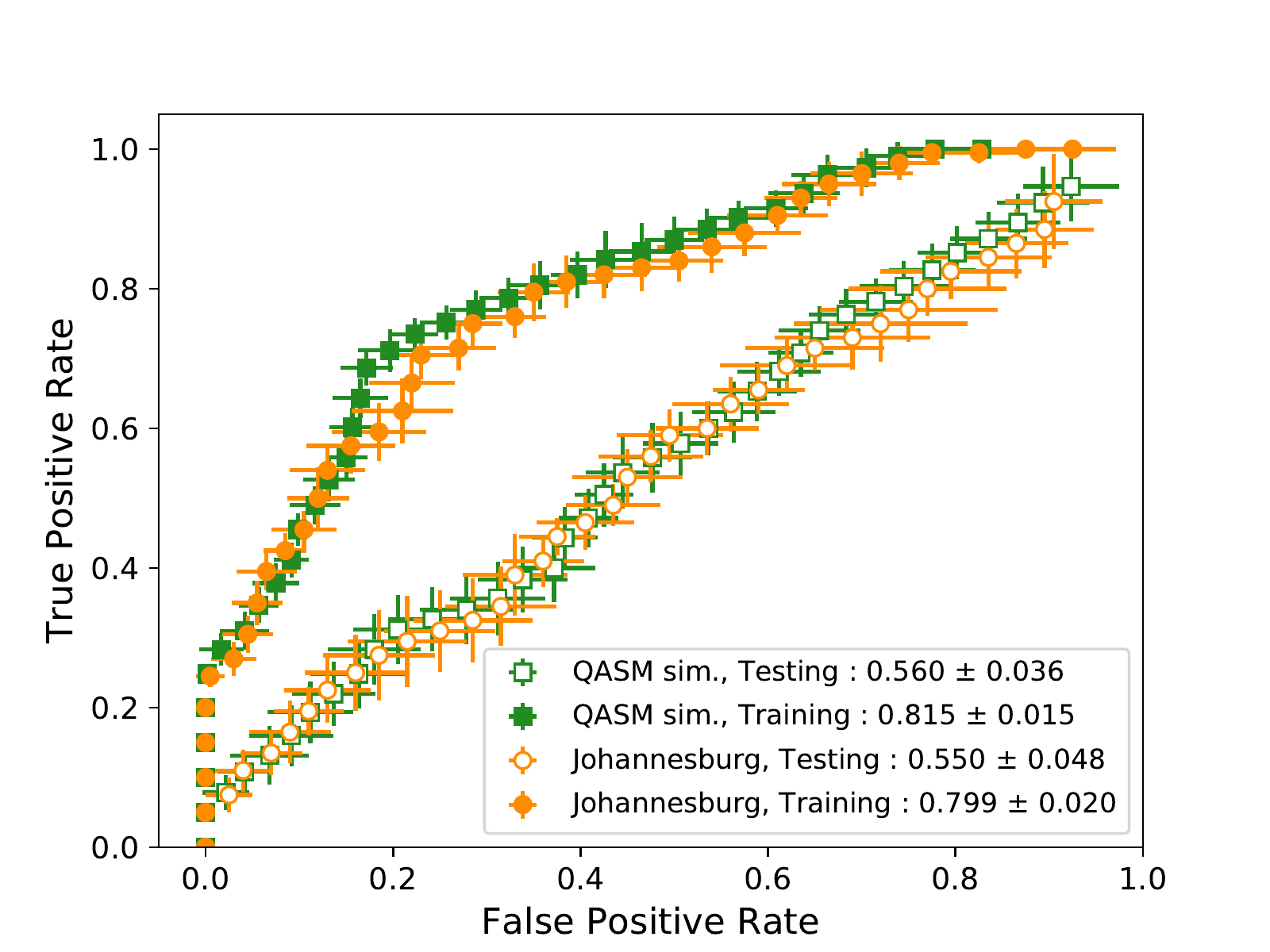}
\caption{ROC curves in the training and testing of the VQC algorithm with $\nvar=3$ and $\nevttr=40$. 
Shown are the ROC curves (averaged over five trials in the training or testing) for quantum computer and QASM simulator.
The size of the markers represents the standard deviation of the trials. The values in the legend give the average 
AUC values and the standard deviations.}
\label{fig:result_roc_vqc} 
\end{figure}

\subsection{Quantum Computer and QASM Simulator}
\label{subsec:qc_result}

The VQC algorithm with $\nvar=3$ has been tested on the 20-qubit IBM Q Network quantum computers 
and the QASM simulator, as explained in Sect.~\ref{subsec:qc}. 
The present study focuses only on the classification accuracy with the real quantum computer. 
Figure~\ref{fig:cost_vqc} shows the values of 
the cost function in the training as a function of \niter for both the quantum computer and the simulator.
For each of the quantum computer and the simulator, 
the training is repeated five times over the same set of events and 
their cost-function values are shown. 
When running the algorithm on the quantum computer, the first three hardware qubits [0, 1, 2] are used~\cite{IBMQConfMap}.
The figure shows that both the quantum computer and the simulator have
reached the minimum values in the cost function after iterating about 50 times. 
However, the cost values for the quantum computer are constantly higher and 
more fluctuating after reaching the minimum values.

\begin{table}
\caption{AUC values in the testing and training for the VQC algorithm running the QASM
simulator. The training condition is fixed to $\nvar=3$ and $\niter=100$ for all \nevttr cases.
The uncertainties correspond to the standard deviations of the average AUC values over the trials.}
\label{tab:vqc_auc_nevt}
\begin{tabular}{lll}
\hline\noalign{\smallskip}
\nevttr ($=\nevtts$) & Testing & Training  \\
\noalign{\smallskip}\hline\noalign{\smallskip}
40     &   $0.555\pm0.032$  &  $0.813\pm0.012$\\
70     &   $0.716\pm0.037$  &  $0.741\pm0.022$\\
100   &   $0.708\pm0.039$  &  $0.761\pm0.025$\\
200   &   $0.812\pm0.012$  &  $0.741\pm0.014$\\
500   &   $0.779\pm0.008$  &  $0.796\pm0.007$\\
1000 &   $0.779\pm0.008$   &  $0.789\pm0.005$\\
\noalign{\smallskip}\hline
\end{tabular}
\end{table}

\begin{figure}
\includegraphics[width=0.53\textwidth]{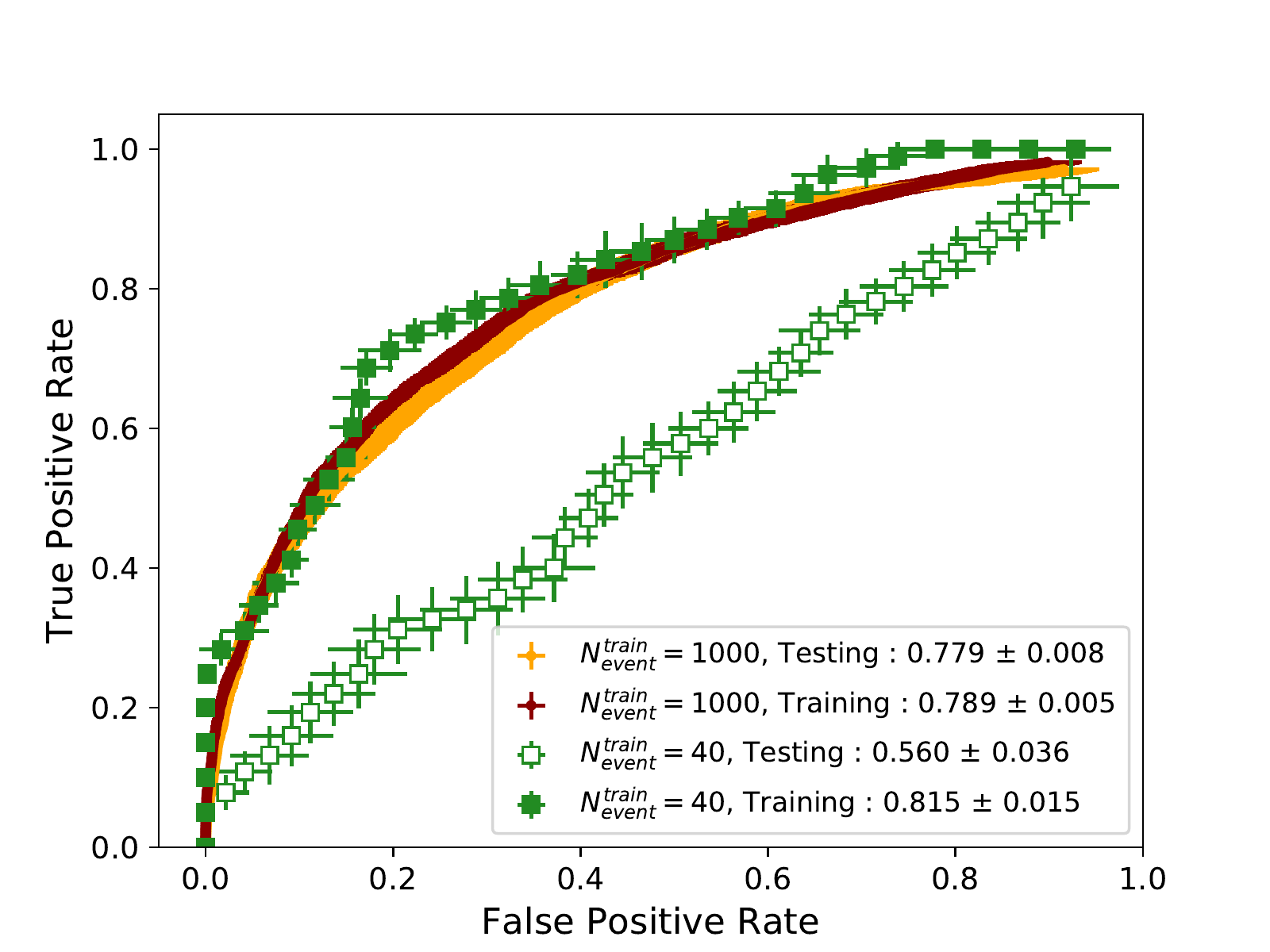}
\caption{ROC curves in the training and testing of the VQC algorithm with $\nevttr=40$ and 1,000
for $\nvar=3$.
Shown are the ROC curves (averaged over five trials in the training or testing) for QASM simulator.
The size of the markers or the band width represent the standard deviation of the trials. 
The values in the legend give the average AUC values and the standard deviations.}
\label{fig:result_roc_vqc_nevt} 
\end{figure}

The ROC curves for the quantum computer and the simulator obtained from the training and testing samples
are shown in Fig.~\ref{fig:result_roc_vqc}, averaged over the five trials of the training or testing.
The AUC values for the testing samples are considerably worse than those for the training ones because 
of the small sample sizes. 
This has been checked by increasing the \nevttr from 40 to 70, 100, 200, 500 and 1,000
for the simulator (Table~\ref{tab:vqc_auc_nevt}). As seen in the table, the over-training largely disappears 
as the sample sizes increase. Figure~\ref{fig:result_roc_vqc_nevt} shows the ROC curves from the simulator for 
the two sample sizes of $\nevttr=40$ and 1,000, confirming that the over-taining is not significant 
for the latter.

The AUC values are consistent between the quantum computer and the simulator within 
the standard deviation (Fig.~\ref{fig:result_roc_vqc}), 
but the simulator results are considered to be systematically better because 
the input samples are identical. 
In Table~\ref{tab:vqc_auc}, the VQC results are compared with the QCL being executed 
at the same condition, i.e, $\nvar=3$, $\nevttr=40$ and $\niter=100$. 
The QCL results vary with the depth of the \uvar circuit (the nominal \ndepthvar is 3), 
but they agree with the VQC results within relatively large uncertainties. 
Summarized in Table~\ref{tab:vqc_auc} are the AUC values and their standard deviations 
in the training of the VQC and QCL algorithms.

\begin{table}
\caption{AUC values in the training for the VQC and QCL algorithms running on quantum computers and 
simulators. The QCL results are given for $\ndepthvar=1$ and 3. The training condition is fixed to 
$\nvar=3$, $\nevttr=40$ and $\niter=100$ for both algorithms.
The uncertainties correspond to the standard deviations of the average AUC values over the trials.}
\label{tab:vqc_auc}
\begin{tabular}{lll}
\hline\noalign{\smallskip}
& Device/Condition & AUC  \\
\noalign{\smallskip}\hline\noalign{\smallskip}
VQC & Johannesburg &  $0.799\pm0.020$\\
 & Boeblingen &  $0.807\pm0.010$\\
 & QASM simulator & $0.815\pm0.015$\\
\noalign{\smallskip}\hline
QCL & Qulacs simulator ($\ndepthvar=1$) &  $0.768\pm0.082$\\
 & Qulacs simulator ($\ndepthvar=3$) &  $0.833\pm0.063$\\
\noalign{\smallskip}\hline
\end{tabular}
\end{table}

\section{Discussion}
\label{sec:discussion}
\subsection{Performance with different QCL models}
As seen in Fig.~\ref{fig:result_auc_all}, the QCL performance stays approximately flat 
in \nevttr and gets slightly worse when increasing the \nvar at fixed \nevttr. 
Since the computational resources needed to explore the QCL model with more variables ($\nvar>\approx10$) or larger
sample sizes ($\nevttr>10$K) are beyond our capacity (Sect.~\ref{subsec:QCL_usage}), 
understanding the behavior and the dependence on the \nvar or \nevttr is a subject for future study.

\begin{figure}
\includegraphics[width=0.51\textwidth]{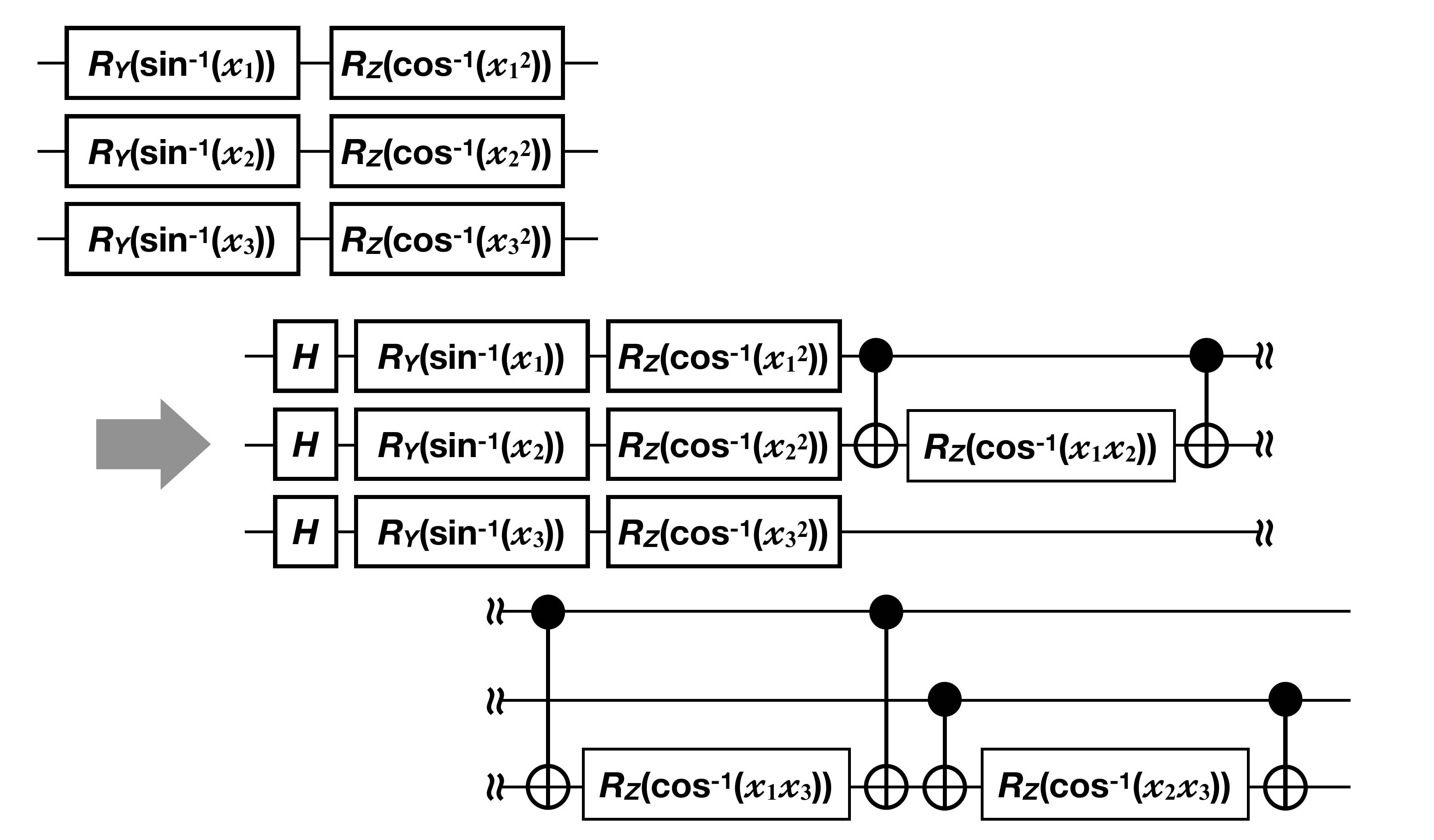}
\caption{Nominal and alternative \uin circuits used in QCL to check impact on the performance.}
\label{fig:qcl_o2uin} 
\end{figure}

To investigate a possibility that the QCL performance could be limited by insufficient flexibility of the circuit used 
(Fig.~\ref{fig:qcl_circuit}),
alternative QCL models with the \uvar circuit of $\ndepthvar=5$ or 7, instead of 3, are tested.
This changes the AUC values
by 1-2\% at most for the \nevttr of 100 or 1,000 events, which is negligible compared to the statistical fluctuation. 
Another type of QCL circuit is also considered by modifying the \uin to include 2-qubit gates for creating entanglement, 
as shown in Fig.~\ref{fig:qcl_o2uin} (as motivated by the Second Order Expansion in VQC; 
see Sect.~\ref{subsec:VQC_perform}).
It turns out that the QCL with the new \uin does not increase the AUC values when the \uvar  
is fixed to the original model with $\ndepthvar=3$ in Fig.~\ref{fig:qcl_circuit}. On the other hand, the new \uin appears to improve 
the performance by 5--10\% with respect to the original \uin when \ndepthvar is set to 1. 
This indicates that a more complex structure in the \uin could help improve the performance when the \uvar is simplified. 
However, the performance of the new \uin with $\ndepthvar =1$ is still considerably worse than the nominal QCL model in Fig.~\ref{fig:qcl_circuit}. 

\subsection{Performance with different VQC models}
\label{subsec:VQC_perform}
The VQC circuit used in this study (Fig.~\ref{fig:vqc_circuit}) is simplified with respect to the one
used in Ref.~\cite{Havlcek2019SupervisedLW}. To examine whether more extended circuits could improve the performance, 
alternative VQC models are tested using the QASM simulator. The first alternative model is the one in which 
the \uphi in Fig.~\ref{fig:vqc_circuit} (FOE) is replaced with the combination of
single- and two-qubit gates of $U_{\phi_{\{k\}}}(\boldsymbol{x})=\exp{(i\phi_{\{k\}}(\boldsymbol{x})Z_k)}$ and
$U_{\phi_{\{l,m\}}}(\boldsymbol{x})=\exp{(i\phi_{\{l,m\}}(\boldsymbol{x})Z_lZ_m)}$ with 
$\phi_{\{l,m\}}(\boldsymbol{x})=(\pi-x_l)(\pi-x_m)$, as used in Ref.~\cite{Havlcek2019SupervisedLW}.
This type of \uphi is referred to as the ``Second Order Expansion" (SOE).
The second alternative model is the one with extended \uin and \uvar gates
by increasing the \ndepthin and \ndepthvar; this model includes the combinations of \ndepthin up to 2 
and \ndepthvar up to 3, separately for the FOE and SOE in \uphi.

Testing these models using the QASM simulator show that 
the AUC values stay almost constant (within at most 2\%) regardless of the \ndepthin or \ndepthvar 
if the \uphi is fixed to either FOE or SOE. But, the performance improves by about 10\% 
when changing the \uphi from FOE to SOE at fixed \ndepthin and \ndepthvar. 
On the other hand, no improvement is observed when testing the SOE with a real quantum computer.
Moreover, the standard deviation of the AUC values becomes significantly larger for the SOE with quantum computer. 
These could be qualitatively understood to be due to increased errors from hardware noise because 
the number of single- and two-qubit gate operations increases by 60\% when switching from the FOE to SOE at 
$\ndepthin=\ndepthvar=1$, therefore the VQC circuit with SOE suffers more from the gate errors.

\subsection{Comparison with DNN model with less number of parameters}
\begin{figure}
\includegraphics[width=0.53\textwidth]{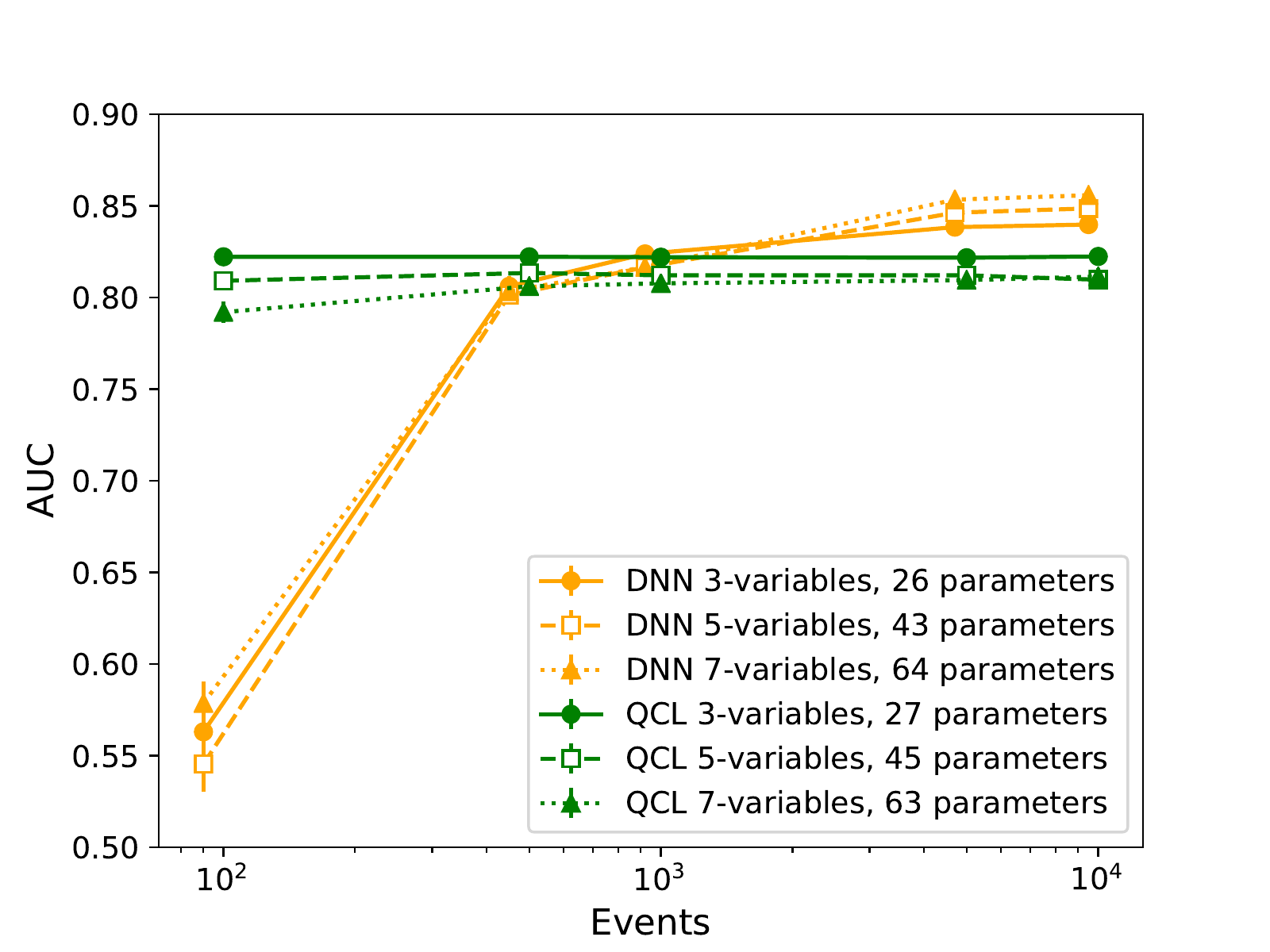}
\caption{Average AUC values (calculated from the test samples) as a function of the training sample size up to $ \nevttr=10,000$
for the new DNN and QCL models with $\nvar=3$ (circles), 5 (squares) and 7 (triangles). 
The error bars represent the standard deviations of the average AUC values. The DNN points are 
slightly shifted horizontally from the nominal \nevttr values to avoid overlapping.}
\label{fig:result_auc_DNN_npar} 
\end{figure}

A characteristic difference between the QCL and DNN algorithms is on the number of trainable parameters (\npar). 
As in Sect.~\ref{subsec:qc_algo}, the \npar is fixed to 27 (45, 63) for the QCL with 3 (5, 7) variable case. 
For the DNN model in Table~\ref{tab:bdt_dnn_param}, the \npar varies with \nevttr as given in Table~\ref{tab:DNN_npar}. 
Typically the \npar of the DNN model is about 6-13 times more than that of the QCL model at 
$\nevttr=100$, and the ratio increases to 75-165 (200-470) at $\nevttr=1,000$ (10,000).
Comparing the two algorithms with a similar number of trainable parameters could give more insight into the QCL performance 
and reveal a potential advantage of the variational quantum approach over the classical method.
A new DNN model is thus constructed to contain only one hidden layer with 5 (6, 7) nodes for 3 (5, 7) variable case,
resulting in the \npar of 26 (43, 64). The rest of the model parameters is identical to that in
Table~\ref{tab:bdt_dnn_param}.
Shown in Fig.~\ref{fig:result_auc_DNN_npar} is the comparison of the AUC values for the new DNN and QCL models
at $\nevttr\leq10,000$. It is indicated from the figure that the QCL can learn more efficiently than 
the simple feed-forward network with the similar number of parameters when the sample size is below 1,000. 
Exploiting this feature in the application to HEP data analysis would be an interesting future subject.

\begin{table}
\caption{Number of trainable parameters used in the DNN model of Table~\ref{tab:bdt_dnn_param}.}
\label{tab:DNN_npar}
\begin{tabular}{llll}
\hline\noalign{\smallskip}
\nevttr & \multicolumn{3}{c}{\npar} \\
& $\nvar=3$ & $\nvar=5$ & $\nvar=7$ \\
\noalign{\smallskip}\hline\noalign{\smallskip}
100         &  353 & 385 & 417 \\
500         &  625 & 657 & 689 \\
1,000      &  4,481 & 4,609 & 4,737 \\
5,000      &  12,801 & 12,929 & 13,057 \\
10,000    &  12,801 & 12,929 & 13,057 \\
50,000    &  50,117 & 50,433 & 50,689 \\
100,000  &  50,117 & 50,433 & 50,689 \\
200,000  &  330,241 & 330,753 & 331,265 \\
500,000  &  330,241 & 330,753 & 331,265 \\ 
\noalign{\smallskip}\hline
\end{tabular}
\end{table}

\subsection{CPU/memory usages for QCL implementation}
\label{subsec:QCL_usage}
The QCL algorithm runs on the Qulacs simulator with cloud Linux servers, as described in Sect.~\ref{subsec:sim}.
Under this condition, we examine how the computational resources scale with the problem size.
For the creation of input quantum states with \uin, 
both CPU time and memory usage grow approximately linearly with \nvar or \nevttr. 
The creation of the variational quantum states with \uvar shows an exponential increase
in CPU time and memory usage with \nvar (i.e, number of qubits) up to $\nvar=12$,
roughly a factor 8 (4) increase in CPU time (memory) by incrementing the \nvar by one.
The overall CPU time is by far dominated by the minimization process with COBYLA. 
It increases linearly with \nevttr but grows exponentially with \nvar, making it impractical to run the algorithm
a sufficient number of times for $\nvar\sim10$ or more. 
The memory usage stays constant over \nvar during the COBYLA minimization process.

\section{Conclusion}
\label{sec:conclusion}
In this paper, we present studies of quantum machine learning for the 
event classification,
commonly used as the application of conventional machine learning techniques to high-energy physics.
The studies focus on the application of variational quantum algorithms using the implementations 
in QCL and VQC, and evaluate the performance in terms of AUC values of the ROC curves.
The QCL performance is compared with the standard classical multi-variate classification 
techniques based on the BDT and DNN, and the VQC performance is tested using the simulator
and real quantum computers. The overall QCL performance is comparable to the standard techniques 
if the problem is restricted to $\nvar\leq7$ and $\nevttr<\sim10,000$. 
The QCL algorithm shows relatively flat AUC values in \nevttr, 
in contrast to the BDT and DNN algorithms, which show that 
the AUC values increase with increasing \nevttr in the considered \nevttr range.
This characteristic QCL behavior could be considered as a possible advantage over
the classical method at small \nevttr where the DNN performance gets considerably worse if the 
number of trainable parameters of the DNN model is constrained to be similar to that of the QCL.

The VQC algorithm has been tested on quantum computers only for a small problem of $\nevttr=40$, but it shows 
that the algorithm does acquire the discrimination power. 
The VQC performances are similar on the simulator and real quantum computer within the measured accuracy.
There is however an indication of increased errors due to hardware noise, that could prevent us from using an extended 
quantum circuit such as the SOE for the encoding of classical input data.
The QCL and VQC algorithms show similar performance 
when they run on the simulators with the same conditions for the \nvar and \nevttr values. 

With a better control of the measurement and gate errors, it is expected that 
the performance of the variational quantum machine learning on quantum computers further improves,
as demonstrated in Ref.~\cite{Havlcek2019SupervisedLW} with the SOE and increased depth of the variational circuit.
Another potentially promising approach, proposed in Ref.~\cite{Lloyd2020QuantumEF}, 
is to train the encoding part of the variational algorithm to carry out a maximally 
separating embedding of classical input data into Hilbert space. This could provide a way to perform 
optimized measurements to distinguish different classes of data with a shallow quantum circuit, 
potentially reducing the impact from hardware errors on the variational part of the algorithm.

\begin{acknowledgements}
We acknowledge the use of IBM Quantum services for this work. The views expressed are those 
of the authors, and do not reflect the official policy or position of IBM or the IBM Quantum team.

We thank Dr. Naoki Kanazawa (IBM Japan), Dr. Tamiya Onodera (IBM Japan) and 
Prof. Hiroshi Imai (The University of Tokyo) for the useful discussion and guidance for 
addressing the technical issues occurred during the studies.

We acknowledge the support from using the dataset provided by the UCI Machine Learning Repository 
in the University of California Irvine, School of Information and Computer Science.

\end{acknowledgements}

\section*{Conflict of interest}
On behalf of all authors, the corresponding author states that there is no conflict of interest.

\section*{Code availability}
The analysis code used in this study is available in 
https://github.com/kterashi/QML\_HEP.

\bibliographystyle{spmpsci_unsort}
\bibliography{QML_paper}   

\begin{thebibliography}{10}
\providecommand{\url}[1]{{#1}}
\providecommand{\urlprefix}{URL }
\expandafter\ifx\csname urlstyle\endcsname\relax
  \providecommand{\doi}[1]{DOI~\discretionary{}{}{}#1}\else
  \providecommand{\doi}{DOI~\discretionary{}{}{}\begingroup
  \urlstyle{rm}\Url}\fi

\bibitem{hepmllivingreview}
{HEP ML Community}: {A Living Review of Machine Learning for Particle Physics}.
\newblock \urlprefix\url{https://iml-wg.github.io/HEPML-LivingReview/}

\bibitem{ApollinariG.:2017ojx}
Apollinari, G., Béjar~Alonso, I., Brüning, O., Fessia, P., Lamont, M., Rossi,
  L., Tavian, L.: {High-Luminosity Large Hadron Collider (HL-LHC)}.
\newblock CERN Yellow Rep. Monogr. \textbf{4}, 1--516 (2017).
\newblock \doi{10.23731/CYRM-2017-004}

\bibitem{Evans:2008zzb}
Evans, L., Bryant, P.: {LHC Machine}.
\newblock JINST \textbf{3}, S08001 (2008).
\newblock \doi{10.1088/1748-0221/3/08/S08001}

\bibitem{Mott:2017xdb}
Mott, A., Job, J., Vlimant, J.R., Lidar, D., Spiropulu, M.: {Solving a Higgs
  optimization problem with quantum annealing for machine learning}.
\newblock Nature \textbf{550}(7676), 375--379 (2017).
\newblock \doi{10.1038/nature24047}

\bibitem{Zlokapa:2019lvv}
Zlokapa, A., Mott, A., Job, J., Vlimant, J.R., Lidar, D., Spiropulu, M.:
  {Quantum adiabatic machine learning with zooming}  (2019)

\bibitem{Shapoval:2019txi}
Shapoval, I., Calafiura, P.: {Quantum Associative Memory in HEP Track Pattern
  Recognition}.
\newblock EPJ Web Conf. \textbf{214}, 01012 (2019).
\newblock \doi{10.1051/epjconf/201921401012}

\bibitem{Bapst:2019llh}
Bapst, F., Bhimji, W., Calafiura, P., Gray, H., Lavrijsen, W., Linder, L.: {A
  pattern recognition algorithm for quantum annealers}.
\newblock Comput. Softw. Big Sci. \textbf{4}, 1 (2019).
\newblock \doi{10.1007/s41781-019-0032-5}

\bibitem{Zlokapa:2019tkn}
Zlokapa, A., Anand, A., Vlimant, J.R., Duarte, J.M., Job, J., Lidar, D.,
  Spiropulu, M.: {Charged Particle Tracking with Quantum Annealing-Inspired
  Optimization}  (2019)

\bibitem{Tuysuz:2020ocw}
Tüysüz, C., Carminati, F., Demirköz, B., Dobos, D., Fracas, F., Novotny, K.,
  Potamianos, K., Vallecorsa, S., Vlimant, J.R.: {Particle Track Reconstruction
  with Quantum Algorithms}.
\newblock p. 09013 (2020).
\newblock \doi{10.1051/epjconf/202024509013}

\bibitem{Das:2019hrw}
Das, S., Wildridge, A.J., Vaidya, S.B., Jung, A.: {Track clustering with a
  quantum annealer for primary vertex reconstruction at hadron colliders}
  (2019)

\bibitem{Wei:2019rqy}
Wei, A.Y., Naik, P., Harrow, A.W., Thaler, J.: {Quantum Algorithms for Jet
  Clustering}.
\newblock Phys. Rev. D \textbf{101}(9), 094015 (2020).
\newblock \doi{10.1103/PhysRevD.101.094015}

\bibitem{Bauer:2019qx}
Provasoli, D., Nachman, B., Bauer, C., de~Jong, W.A.: A quantum algorithm to
  efficiently sample from interfering binary trees.
\newblock Quantum Science and Technology \textbf{5}(3), 035004 (2020).
\newblock \doi{10.1088/2058-9565/ab8359}

\bibitem{Bauer:2019qxa}
Bauer, C.W., De~Jong, W.A., Nachman, B., Provasoli, D.: {A quantum algorithm
  for high energy physics simulations}  (2019)

\bibitem{Cormier:2019kcq}
Cormier, K., Di~Sipio, R., Wittek, P.: {Unfolding measurement distributions via
  quantum annealing}.
\newblock JHEP \textbf{11}, 128 (2019).
\newblock \doi{10.1007/JHEP11(2019)128}

\bibitem{Bauer:2019uf}
Bauer, C.W., De~Jong, W.A., Nachman, B., Urbanek, M.: {Unfolding quantum
  computer readout noise}.
\newblock npj Quantum Inf. \textbf{6}, 84 (2020).
\newblock \doi{10.1038/s41534-020-00309-7}

\bibitem{Preskill2018quantumcomputingin}
Preskill, J.: Quantum {C}omputing in the {NISQ} era and beyond.
\newblock {Quantum} \textbf{2}, 79 (2018).
\newblock \doi{10.22331/q-2018-08-06-79}.
\newblock \urlprefix\url{https://doi.org/10.22331/q-2018-08-06-79}

\bibitem{Johnson2011Quantum}
Johnson, M.W., Amin, M.H.S., Gildert, S., Lanting, T., Hamze, F., Dickson, N.,
  Harris, R., Berkley, A.J., Johansson, J., Bunyk, P., Chapple, E.M., Enderud,
  C., Hilton, J.P., Karimi, K., Ladizinsky, E., Ladizinsky, N., Oh, T.,
  Perminov, I., Rich, C., Thom, M.C., Tolkacheva, E., Truncik, C.J.S.,
  Uchaikin, S., Wang, J., Wilson, B., Rose, G.: Quantum annealing with
  manufactured spins.
\newblock Nature \textbf{473}(7346), 194--198 (2011).
\newblock \doi{10.1038/nature10012}.
\newblock \urlprefix\url{http://dx.doi.org/10.1038/nature10012}

\bibitem{Peruzzo2014vqe}
Peruzzo, A., McClean, J., Shadbolt, P., Yung, M.H., Zhou, X.Q., Love, P.,
  Aspuru-Guzik, A., O'Brien, J.: A variational eigenvalue solver on a photonic
  quantum processor.
\newblock Nature Communications \textbf{5} (2014).
\newblock \doi{10.1038/ncomms5213}

\bibitem{Mitarai_2018}
Mitarai, K., Negoro, M., Kitagawa, M., Fujii, K.: Quantum circuit learning.
\newblock Physical Review A \textbf{98}(3) (2018).
\newblock \doi{10.1103/physreva.98.032309}.
\newblock \urlprefix\url{http://dx.doi.org/10.1103/PhysRevA.98.032309}

\bibitem{Havlcek2019SupervisedLW}
Havl{\'i}cek, V., C{\'o}rcoles, A.D., Temme, K., Harrow, A.W., Kandala, A.,
  Chow, J.M., Gambetta, J.M.: Supervised learning with quantum-enhanced feature
  spaces.
\newblock Nature \textbf{567}, 209--212 (2019).
\newblock \doi{10.1038/s41586-019-0980-2}

\bibitem{scikit-learn}
Pedregosa, F., Varoquaux, G., Gramfort, A., Michel, V., Thirion, B., Grisel,
  O., Blondel, M., Prettenhofer, P., Weiss, R., Dubourg, V., Vanderplas, J.,
  Passos, A., Cournapeau, D., Brucher, M., Perrot, M., Duchesnay, E.:
  Scikit-learn: Machine learning in {P}ython.
\newblock Journal of Machine Learning Research \textbf{12}, 2825--2830 (2011)

\bibitem{Speckmayer_2010}
Speckmayer, P., Höcker, A., Stelzer, J., Voss, H.: The toolkit for
  multivariate data analysis, {TMVA} 4.
\newblock Journal of Physics: Conference Series \textbf{219}(3), 032057 (2010).
\newblock \doi{10.1088/1742-6596/219/3/032057}.
\newblock \urlprefix\url{https://doi.org/10.1088/1742-6596/219/3/032057}

\bibitem{IBMQ}
{IBM} {Q} {N}etwork.
\newblock
  \urlprefix\url{https://www.ibm.com/quantum-computing/network/overview/}

\bibitem{Dua:2019}
Dua, D., Graff, C.: {UCI} machine learning repository (2017).
\newblock \urlprefix\url{http://archive.ics.uci.edu/ml}

\bibitem{Baldi:2014kfa}
Baldi, P., Sadowski, P., Whiteson, D.: {Searching for Exotic Particles in
  High-Energy Physics with Deep Learning}.
\newblock Nature Commun. \textbf{5}, 4308 (2014).
\newblock \doi{10.1038/ncomms5308}

\bibitem{Qulacs}
Qulacs.
\newblock \urlprefix\url{http://qulacs.org/index.html}

\bibitem{Qiskit}
Abraham, H., Akhalwaya, I.Y., Aleksandrowicz, G., Alexander, T., Alexandrowics,
  G., Arbel, E., Asfaw, A., Azaustre, C., AzizNgoueya, Barkoutsos, P., Barron,
  G., Bello, L., Ben-Haim, Y., Bevenius, D., Bishop, L.S., Bosch, S., Bucher,
  D., CZ, Cabrera, F., Calpin, P., Capelluto, L., Carballo, J., Carrascal, G.,
  Chen, A., Chen, C.F., Chen, R., Chow, J.M., Claus, C., Clauss, C., Cross,
  A.J., Cross, A.W., Cross, S., Cruz-Benito, J., Cryoris, Culver, C.,
  C{\'o}rcoles-Gonzales, A.D., Dague, S., Dartiailh, M., DavideFrr, Davila,
  A.R., Ding, D., Drechsler, E., Drew, Dumitrescu, E., Dumon, K., Duran, I.,
  Eastman, E., Eendebak, P., Egger, D., Everitt, M., Fern{\'a}ndez, P.M.,
  Fern{\'a}ndez, P.M., Ferrera, A.H., Frisch, A., Fuhrer, A., GEORGE, M.,
  GOULD, I., Gacon, J., Gadi, Gago, B.G., Gambetta, J.M., Garcia, L., Garion,
  S., Gawel-Kus, Gomez-Mosquera, J., de~la Puente~Gonz{\'a}lez, S., Greenberg,
  D., Grinko, D., Guan, W., Gunnels, J.A., Haide, I., Hamamura, I., Havlicek,
  V., Hellmers, J., Herok, {\L}., Hillmich, S., Horii, H., Howington, C., Hu,
  S., Hu, W., Imai, H., Imamichi, T., Ishizaki, K., Iten, R., Itoko, T.,
  Javadi-Abhari, A., Jessica, Johns, K., Kanazawa, N., Kang-Bae, Karazeev, A.,
  Kassebaum, P., Knabberjoe, Kovyrshin, A., Krishnan, V., Krsulich, K., Kus,
  G., LaRose, R., Lambert, R., Latone, J., Lawrence, S., Liu, D., Liu, P., Mac,
  P.B.Z., Maeng, Y., Malyshev, A., Marecek, J., Marques, M., Mathews, D.,
  Matsuo, A., McClure, D.T., McGarry, C., McKay, D., Meesala, S., Mezzacapo,
  A., Midha, R., Minev, Z., Mooring, M.D., Morales, R., Moran, N., Murali, P.,
  M{\"u}ggenburg, J., Nadlinger, D., Nannicini, G., Nation, P., Naveh, Y.,
  Nick-Singstock, Niroula, P., Norlen, H., O'Riordan, L.J., Ogunbayo, O.,
  Ollitrault, P., Oud, S., Padilha, D., Paik, H., Perriello, S., Phan, A.,
  Pistoia, M., Pozas-iKerstjens, A., Prutyanov, V., Puzzuoli, D., P{\'e}rez,
  J., Quintiii, Raymond, R., Redondo, R.M.C., Reuter, M., Rodr{\'\i}guez, D.M.,
  Ryu, M., SAPV, T., SamFerracin, Sandberg, M., Sathaye, N., Schmitt, B.,
  Schnabel, C., Scholten, T.L., Schoute, E., Sertage, I.F., Shammah, N., Shi,
  Y., Silva, A., Siraichi, Y., Sitdikov, I., Sivarajah, S., Smolin, J.A.,
  Soeken, M., Steenken, D., Stypulkoski, M., Takahashi, H., Taylor, C.,
  Taylour, P., Thomas, S., Tillet, M., Tod, M., de~la Torre, E., Trabing, K.,
  Treinish, M., TrishaPe, Turner, W., Vaknin, Y., Valcarce, C.R., Varchon, F.,
  Vogt-Lee, D., Vuillot, C., Weaver, J., Wieczorek, R., Wildstrom, J.A., Wille,
  R., Winston, E., Woehr, J.J., Woerner, S., Woo, R., Wood, C.J., Wood, R.,
  Wood, S., Wootton, J., Yeralin, D., Yu, J., Zachow, C., Zdanski, L., Zoufalc,
  anedumla, azulehner, bcamorrison, brandhsn, chlorophyll zz, dime10, drholmie,
  elfrocampeador, faisaldebouni, fanizzamarco, gruu, kanejess, klinvill,
  kurarrr, lerongil, ma5x, merav aharoni, mrossinek, neupat, ordmoj,
  sethmerkel, strickroman, sumitpuri, tigerjack, toural, willhbang, yang.luh,
  yotamvakninibm: Qiskit: An open-source framework for quantum computing
  (2019).
\newblock \doi{10.5281/zenodo.2562110}

\bibitem{johannesburg}
ibmq\_johannesburg, {IBM} {Quantum} team. {Retrieved} from
  https://quantum-computing.ibm.com (2020).
\newblock
  \urlprefix\url{https://quantum-computing.ibm.com/docs/cloud/backends/systems/}

\bibitem{boeblingen}
ibmq\_boeblingen, {IBM} {Quantum} team. {Retrieved} from
  https://quantum-computing.ibm.com (2020).
\newblock
  \urlprefix\url{https://quantum-computing.ibm.com/docs/cloud/backends/systems/}

\bibitem{IBMQConfMap}
{IBM} {Q} system configuration maps.
\newblock
  \urlprefix\url{https://www.ibm.com/blogs/research/2019/09/quantum-computation-center/}

\bibitem{Lloyd2020QuantumEF}
Lloyd, S., Schuld, M., Ijaz, A., Izaac, J.A., Killoran, N.: Quantum embeddings
  for machine learning  (2020)

\end{thebibliography}

\end{document}